\documentclass[12pt,english]{iopart}
\usepackage{iopams}
\usepackage{mathptmx}
\usepackage{graphicx}
\usepackage{amssymb}
\usepackage{babel}
\usepackage[usenames,dvipsnames]{color}

\usepackage[numbers,merge,square,sort&compress]{natbib}

\usepackage{color}
\definecolor{dred}{rgb}{.6,.0,0.}
\definecolor{dblue}{rgb}{.0,.0,0.6}

\begin{document}

\title[Non-equilibrium dynamics of the Dicke model]{Non-equilibrium dynamics of the Dicke model for mesoscopic aggregates: signatures of superradiance}

\author{Sebastian Fuchs$^{1,2,3}$, Joachim Ankerhold$^1$, Miles Blencowe$^3$,  Bj\"orn Kubala$^1$}
\address{$^1$ Institute for Complex Quantum Systems, Ulm University, 89069 Ulm, Germany}
\address{$^2$ Department of Physics and Astronomy, Northwestern University, Evanston, Illinois 60208, USA}
\address{$^3$ Department of Physics and Astronomy, Dartmouth College, Hanover, New Hampshire 03755, USA}

\date{\today}

\begin{abstract}
In the dissipative quantum dynamics of a mesoscopic aggregate of excited two level systems (atoms) coupled to a single resonance mode of
 a cavity, two physical phenomena associated with superradiance appear. A pronounced emission peak on short time scales is related 
 to the known superradiant burst of excited atoms cooperatively radiating into free space. It is followed by relaxation to a stationary state 
 of the composite system such that for sufficiently large atom-field coupling a strongly correlated state emerges. The crossover to this state 
 can be interpreted as the precursor of the transition from normal to superradiant phase in the thermodynamic limit of the Dicke Hamiltonian.
Motivated by recent experimental activities, these features are investigated in detail for a mesoscopic number of atoms and a cavity 
embedded in a dissipative medium described by  a damped Dicke model. We identify observables and characteristics of the quantum dynamics on shorter time scales which allow to clearly distinguish weakly correlated from strongly correlated Dicke physics.

\end{abstract}


%
%
%
%
%
%


\setcounter{tocdepth}{2}
\tableofcontents

\maketitle

\section{Introduction}

The Dicke model, originally developed to describe a large number $N$ of two-level atoms interacting with a single-mode radiation field \cite{dicke:1954, gross:1982}, has regained substantial interest in the last decade \cite{brandes:2005,baumann:2010,*brennecke:2013,*hamner:2014,*baden:2014,*kessler:2014,kindler:2014,mlynek:2014}.
Recent experimental realizations of Dicke physics have utilized various degrees of freedom of cold atom clouds, see e.g.\ \cite{baumann:2010,*brennecke:2013,*hamner:2014,*baden:2014,*kessler:2014,kindler:2014,ritsch:2013}, and implementations with solid state devices have also been discussed \cite{nataf:2010,*viehmann:2011}.

In this context, theory has contributed to a variety of aspects of the model with most of the work basically focusing on two issues, namely, the dissipative \emph{dynamics} in the semiclassical limit of a large collective spin consisting of the individual atomic levels, and properties at and close to the \emph{equilibrium} state in the thermodynamic limit of $N\to \infty$ and vanishing coupling to a dissipative medium. Both regimes display characteristic phenomena associated with the term superradiance: on a transient time scale, the cooperativity of the atoms may lead to superradiant light emission (burst) while thermodynamically a phase transition occurs when the atom-field coupling exceeds a certain threshold. The latter threshold captures the changeover from a regime where energy eigenstates are product states of the atoms and the field mode, to a regime where collective behavior associated with strong atom-field correlations appears (superradiant phase).

What has gained much less attention, is first the relation between these two facets, the superradiant burst and the phase transition, with respect to atom-field correlations. This issue is not only relevant experimentally \cite{kindler:2014}, but also for the analysis of non-equilibrium phase transitions \cite{polkovnikov:2011}. Second, the impact of dissipation and finite size effects on the 
emergence of a superradiant phase have not been fully understood yet, despite their important role for mesoscopic aggregates. 
The purpose of this work is to contribute towards closing this gap.  In particular, we will focus on the non-equilibrium dynamics of the Dicke model 
starting initially  far from equilibrium for a moderate number of atoms (mesoscopic regime).

In fact, experimental progress for tailored superconducting circuits allows to integrate multiple artificial atoms in the form of Cooper pair boxes into micro-cavities. To maintain sufficiently long coherence times and tunability, the number of two-level systems in these devices must be kept moderate and far below the number of atoms in a cold atomic cloud. Furthermore, environmental degrees of freedom in the form of heat baths are always present in these solid state structures. Nevertheless,  one may assume that signatures of Dicke physics are observable \cite{mlynek:2014}, at least in certain ranges of time and parameter space. Another line of research combines atomic ensembles with superconducting cavities to benefit from the long coherence times of the former and the fast tunability of the latter entities \cite{bernon:2013,*bothner:2013,*brune:2014}. In this case, dissipation on the solid state side, i.e.\ the finite photon lifetime in the micro-cavity, is the dominant mechanism and must be taken into account.  

 These new composite systems in turn trigger the more fundamental question, namely,  to what extent the Dicke model is realized in these aggregates at all. For example, the atom-field interaction may include higher order terms in the field operator not captured in the model \cite{nataf:2010,viehmann:2011} or couplings to residual degrees of freedom may be present (e.g.\ trapping loss of atoms, charge-background fluctuations in solid state systems, etc.). The Dicke model may be then justified at least on transient time scales which are short compared to typical relaxation times of the full composite system. Consequently, it may be extremely difficult, if not impossible, to monitor 
 Dicke dynamics on long time scales and towards relaxation to a collective thermodynamic ground state. Here, we thus also address the question whether within a mesoscopic Dicke aggregate, signatures of the true phase transition are encapsulated in superradiant-type phenomena on transient time scales and in the presence of dissipation.

The paper is organized as follows. We start in section~\ref{SR_facets} with a brief discussion of the two characteristic phenomena, namely, the superradiant burst and the superradiant phase and then introduce the dissipative Dicke model explored in the sequel. In section~\ref{cavoccu}, 
the dynamics of the photon population in the field mode (cavity) is studied in detail within the full model which is contrasted with its often used form based on a rotating wave approximation. Scaling properties of the initial radiation peak are analyzed.
The issue of how to understand and monitor experimentally possible collective dynamics of atoms and cavity is addressed in section~\ref{monitor}.

\section{Two facets of superradiance\label{SR_facets}}
 To set the stage, we briefly recap in this section the characteristics of the two well-known facets of superradiance  and discuss how we envision their interplay in the proposed scenario.

\subsection{Light emission into free space: The superradiant burst}
A single excited atom in free space will radiate its energy isotropically in an exponential decay process. If a large number of excited atoms are brought into close proximity their decay is not described by adding the light emission rates from all the atoms independently. Instead a cooperative emission process of all atoms is established, which results in a strong, short anisotropic burst of light. This \emph{superradiant free-space burst}, predicted by Dicke in 1954 \cite{dicke:1954} and first
experimentally observed by Skribanovitz \cite{skribanowitz:1973}, has since been extensively studied and its characteristic features are well understood (see e.g.~\cite{gross:1982} for a review).

For the purpose of this paper, we will consider as defining signature of superradiance the crossover from an exponential decay for a single atom to a peak in the time-dependence of the emission rate. This peak gradually becomes more pronounced when the effective number of atoms, $N$, cooperating in the emission process increases (roughly speaking, $N$ is the number of atoms within a volume, $\lambda_\nu^3$, where $\lambda_\nu$ is the wavelength of the emitted light). Due to the cooperative nature of the light emission, the peak's height grows as $N^2$ while its temporal width shrinks as $N^{-1}$, since the integrated emission is proportional to the total initial excitation energy and, hence, to $N$.

The simplest theoretical description of such burst features relies on a rate model. The radiation of $N$ excited indistinguishable two-level atoms is calculated by determining decay rates between the $N+1$ levels of a spin\--$N/2$ system (which results from adding the (pseudo)spins of two-level atoms, see below) from the corresponding dipole matrix elements \cite{gross:1982}.

\subsection{Thermodynamics: The superradiant phase transition}

\begin{figure}[t]
\centerline{\includegraphics[width=0.95\columnwidth]{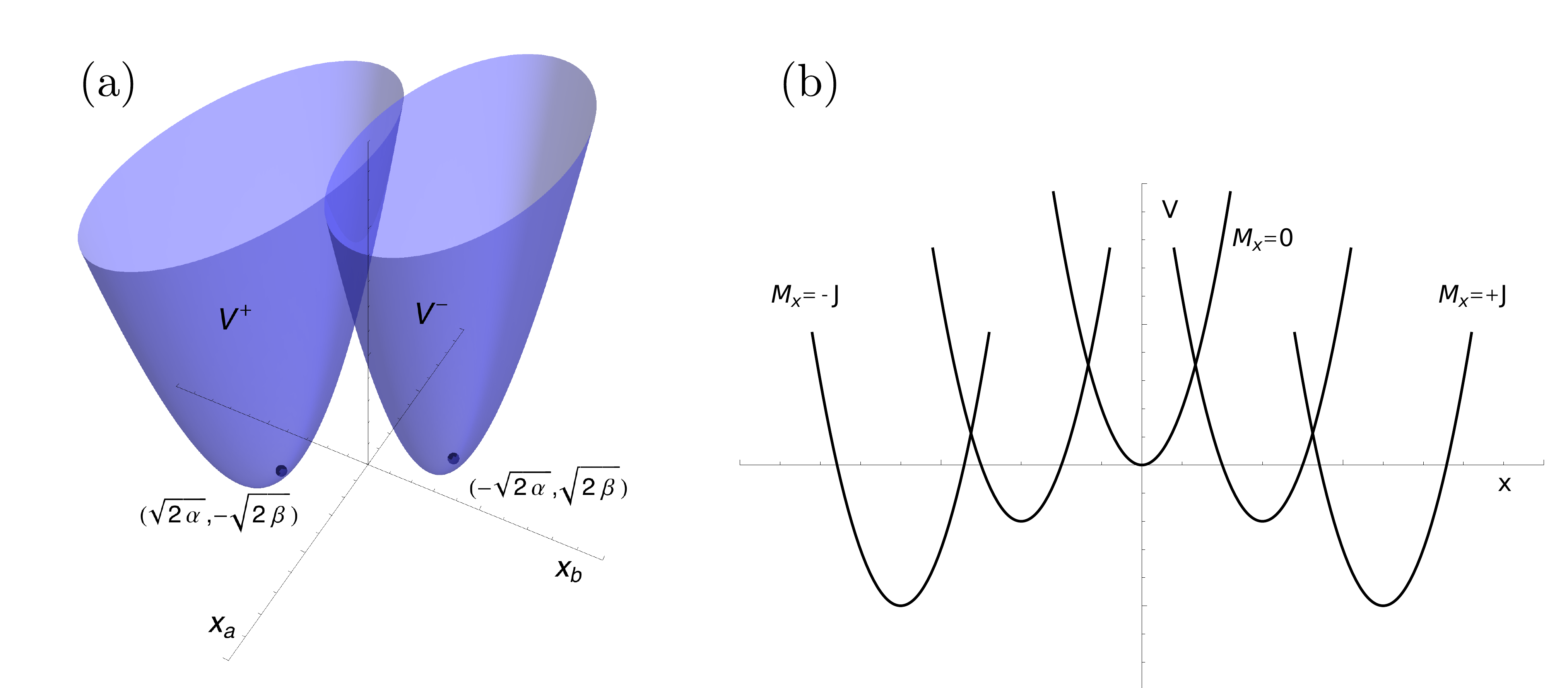}}
\caption{Visualizations of the potential landscape of the Dicke Hamiltonian.\\
(a) The Holstein-Primakoff approach maps cavity and spin degree of freedom to (displaced) bosonic operators (cf.~\eref{HP1} and \eref{HP2}). Above a critical coupling strength two equivalent potentials $V^\pm$ quadratic in the corresponding (dimensionless) position variables $X_{a/b}$ are found, which describe the system close to two mean field solutions $(X_a,\,X_b)=(\pm\sqrt{2\alpha}, \mp \sqrt{2\beta})$, see \eref{HP_meanvalues}.\\
(b) In a $\hat{J}^x$ eigenbasis the Dicke Hamiltonian can be visualized as shifted parabolas for the various $M_x$ eigenvalues, that are coupled by a $\hat{J}^z$  term, see \eref{eq:Dicke-Hamiltonian-Larson}.}
\label{Potential_fig}
\end{figure}

\subsubsection{The Dicke Hamiltonian}
A collection of $N$ identical atoms in a single mode cavity can be described by the Dicke Hamiltonian \cite{dicke:1954,gross:1982,brandes:2005},
\begin{equation}
\label{eq:Dicke-Hamiltonian}
\hat H _{\mathrm{Dicke}} = \omega \hat a ^{\dagger} \hat a + \epsilon \hat J ^z + \frac{\lambda}{\sqrt{N}} \left( \hat a ^{\dagger} + \hat a \right) \left( \hat{J}^+ + \hat{J}^- \right)\;,
\end{equation}
where the atoms are modeled as two-level systems with identical level spacing $\epsilon$ for which collective ("large") spin operators are introduced, $\hat{J}^z = \frac{1}{2} \sum^{N} _{j=1} {\hat \sigma ^z _j}, \; \hat{J}^{\pm} = \sum^{N} _{j=1} {\hat \sigma ^\pm_j}\ , \ \hbar=1$. The collective spin couples to a common, single electromagnetic mode of frequency $\omega$ in the cavity described by bosonic operators $\hat{a}, \hat{a}^\dagger$ with effective coupling strength $\lambda/\sqrt{N}$. The latter one is conveniently introduced when the density of atoms per unit volume is fixed since then the bare coupling $\lambda$ between any single atom and the atomic mode is effectively reduced with a growing number of atoms. If instead in a given experimental situation the actual density of atoms can be increased, $\lambda$ grows accordingly.

The above modeling (see~\cite{brandes:2005} for a recent review) relies on the assumption that all atoms couple to the mode with equal strength (cf. e.g.,~\cite{braun:2011,zou:2014} for inhomogeneous effects) and that a loss of atoms from the collective state (e.g.\ due to trapping losses or external noise sources, see \cite{henschel:2010}) can be neglected on the timescales of interest. Accordingly, the collective spin operators in (\ref{eq:Dicke-Hamiltonian}) couple only states within one $J$ manifold of the so-called Dicke states $| J,M \rangle$. For example, these are states with maximum angular momentum $J=N/2$ when the atomic system is initialized in the atomic ground state $| J=N/2,M=-J \rangle$ or the maximally inverted state  $| J=N/2,M=+J \rangle$. The validity of the Dicke Hamiltonian for various experimental scenarios with (artificial) atoms has been widely discussed recently \citep{zakowicz:1975,*knight:1978,*bialynicki-birula:1979,*keeling:2007,*viehmann:2011,*vukics:2012,*vukics:2014}.
Within a rotating wave approximation (RWA), where only the coupling terms $\hat a ^{\dagger} \hat J ^- + \hat a \hat J ^+$ are kept, the Hamiltonian conserves  in addition to the spin magnitude $J$  also the total number of photonic and atomic excitations $\hat N _{\mathrm{ex}} = \hat a ^{\dagger} \hat a + \hat J ^z + J$. The full (non-RWA) Hamiltonian (\ref{eq:Dicke-Hamiltonian}) only conserves the parity $\hat \Pi = \exp{(\rmi \pi \hat N _{\mathrm{ex}})}$.

\subsubsection{Phase Transition}
Considering the ground state of the Dicke Hamiltonian, it is immediately obvious, that there is a competition between the coupling term and the terms describing the energy of the uncoupled cavity and spin system, respectively. This competition depends on the strength of the coupling $\lambda$ compared to the excitation energy for a cavity photon $\omega$ or a single spin excitation $\epsilon$. For weak coupling, the ground state of the composite system lies close to a product state of the individual ground states of isolated cavity and isolated spin, i.e.\,
$|\mathrm{GS}_{\lambda=0}\rangle:= | n=0,\,M =-N/2\rangle$. Of course, this state is a common eigenstate of the cavity number operator $\hat{n}= \hat a ^{\dagger} \hat a$ and $\hat{J}^z$ with $\hat{n}\, |\mathrm{GS}_{\lambda=0}\rangle =0$
and $\hat{J}^z\, |\mathrm{GS}_{\lambda=0}\rangle = -(N/2)\,   |\mathrm{GS}_{\lambda=0}\rangle$.

In the opposite regime of strong coupling, one has a finite expectation value for the operator $(\hat a ^{\dagger} + \hat a) (\hat{J}^++\hat{J}^-)$ associated with a strongly correlated ground state with non-zero photon occupation  and a finite number of spin excitations. In fact, in the thermodynamic limit $N\rightarrow \infty$, the Dicke Hamiltonian was shown \cite{hepp_annals:1973,*wang_and_hioe:1973,hepp:1973} to exhibit a phase transition between a superradiant phase with macroscopic occupations in the field and the atoms and a normal phase without excitations at zero temperature.  This transition occurs at a critical coupling strength $\lambda_\mathrm{c}=\sqrt{\omega\epsilon}/2$ \cite{hepp:1973} and at $\lambda^\mathrm{RWA}_\mathrm{c}=\sqrt{\omega\epsilon}$ for the RWA version of the Dicke Hamiltonian \cite{hepp_annals:1973,*wang_and_hioe:1973}.

The most intuitive picture for the emergence of this transition and the nature of the respective ground states is provided by the Holstein-Primakoff (HP) approach \cite{emary:2003_2}. The HP treatment of the Dicke Hamiltonian essentially consists of (see~\cite{emary:2003_2} for details):\\
(i) Introducing bosonic operators $\hat{b},\hat{b}^\dagger$ to replace the large spin operator in a HP transformation
\begin{equation}
\label{HP1}
\hat J ^z = \hat b ^{\dagger} \hat b - J, \; \hat J ^+ = \hat b ^{\dagger} \sqrt{2J - \hat b ^{\dagger} \hat b}, \; \hat J ^- = \sqrt{2J - \hat b ^{\dagger} \hat b} \hat b\; .
\end{equation}
(ii) Introducing displaced operators
\begin{equation}
\label{HP2}
\hat a \rightarrow \hat c \pm \sqrt{\alpha}, \quad \hat b \rightarrow \hat d \mp \sqrt{\beta},
\end{equation}
with real-valued parameters $\alpha,\beta$ to be determined self-consistently. In the superradiant phase they are assumed to be of order ${\cal{O}}(N)$, corresponding to a macroscopic number of excitations.\\
(iii)  Performing the HP approximation, a lowest order expansion in the scaled number of excitations $\hat{b
}^\dagger \hat{b}/N$ assuming only small fluctuations around the mean field solutions. This leads to a Hamiltonian bilinear in the bosonic operators $\hat c, \hat d$ , while linear terms are eliminated by the self-consistent choice for the displacements (Bogoliubov transformation).

It then turns out that for $\lambda < \lambda_\mathrm{c}$ there is only a trivial solution $\alpha=0=\beta$, while for $\lambda > \lambda_\mathrm{c}$ the displacements
\begin{equation}
\sqrt{\alpha} = \frac{2 \lambda}{\omega} \sqrt{\frac{J}{2} [ 1- (\lambda_\mathrm{c}/\lambda)^4 ]}\ ,\
\sqrt{\beta} = \sqrt{J \left[ 1 -  (\lambda_\mathrm{c}/\lambda)^2  \right]}
\end{equation}
indeed correspond to macroscopic expectation values
\begin{equation}
\label{HP_meanvalues}
\frac{\langle \hat a ^{\dagger} \hat a \rangle _{\mathrm{SP}}}{J} = \frac{2 \left( \lambda ^4 - \lambda ^4 _\mathrm{c} \right)}{\left( \omega \lambda \right)^2}
\; ; \quad
\frac{\langle \hat J ^z \rangle _{\mathrm{SP}}}{J} = \frac{\beta}{J} - 1 = - \frac{\lambda ^2 _\mathrm{c}}{\lambda ^2}\;
\end{equation}
with $\langle \cdot\rangle_{\rm SP}$ denoting expectation values in the superradiant phase.

The respective Hamiltonians above and below the transition can also be formulated in terms of two sets of conjugate operators 
$X_{a/b},\,P_{a/b}$, i.e.\ positions and momenta for the respective operators of cavity and atomic degrees of freedom. The position degrees of freedom $X_{a/b}$ are coupled via
bilinear potential terms which provides a very intuitive understanding of the emergence of a collective ground state. Namely, the principal axes are rotated with respect to the original cavity and atomic direction such that the potential surface becomes unstable in one direction when $\lambda$ exceeds the critical value $\lambda_\mathrm{c}$. Correspondingly, the HP approximation based on the assumption that fluctuations remain small around the fixed points $\alpha=\beta=0$ fails. It only applies in the normal phase 
$\lambda<\lambda_\mathrm{c}$ sufficiently away from the phase transition, where then the Dicke model is well described by a 
two-dimensional harmonic oscillator potential centered around $X_a=X_b=0$. Instead, in the superradiant phase  $\lambda>\lambda_\mathrm{c}$
 the HP approach leads to two quadratic potentials $V^\pm$ centered around either one of the solutions, $(X_a,\,X_b)=(\pm\sqrt{2\alpha}, \mp \sqrt{2\beta})$ [see \fref{Potential_fig}(a)].

In this latter regime, eigenstates of the model which respect the parity symmetry of the original Dicke Hamiltonian, are constructed from (anti)symmetric superpositions of the individual states of  $H^\pm$. In the thermodynamic limit, the HP approximation becomes exact and corresponding energy eigenvalues are double degenerate. This degeneracy is lifted for any finite number $N$ of atoms by quantum tunneling between the wells of $V^\pm$, a non-locality which cannot be captured by the HP approach. Moreover, the Hamiltonians derived within the HP approach are only suited to study the low energy sector 
of superradiant and normal phase, respectively. They definitely fail to describe the non-equilibrium dynamics when the system is initialized far from equilibrium, e.g.\ in a highly excited spin state, $|n=0, M=+N/2\rangle$.

To go beyond these limitations, an alternative formulation of the Dicke Hamiltonian \cite{larson:2007} may offer some insight. For this purpose, one introduces collective operators $\hat{\underline{a}}^{\dagger},  \hat{\underline{a}} $ with $\hat{\underline{a}}=\hat{a}+2\lambda/(\omega\sqrt{N})\,\hat{J}^x$  so that
\begin{equation}
\label{eq:Dicke-Hamiltonian-Larson}
\hat{H} _{\mathrm{Dicke}} = \omega \hat{\underline{a}}^{\dagger}  \hat{\underline{a}} - \frac{4 \lambda^2}{N\omega} \left(\hat{J}^x\right)^2  + \epsilon \hat J ^z \; .
\end{equation}
In a $\hat{J}^x$ eigenbasis, Dicke physics can then be visualized as governed by a set of shifted parabolas with nearest neighbor coupling
described by the $\epsilon \hat{J}^z$ term (see \fref{Potential_fig}). Starting for instance from the highly excited  $\hat{J}^z$ eigenstate $|n=0, M=+N/2\rangle$, the corresponding quantum dynamics includes  Landau-Zener-type of transitions through a multitude of avoided level crossings. Details of the corresponding time evolution will be discussed elsewhere.

\subsection{Non-equilibrium dynamics of the  Dicke model}

The above discussion brings us to the question concerning to what extent the two facets of superradiant physics can appear 
within a single experimental setup.
Here we argue that in a minimal setting it suffices to add dissipation in the field mode and then to consider the relaxation dynamics of the compound from a maximally excited atomic state.

In the regime of very strong damping, where the level broadening in the cavity mode by far exceeds the coupling energy between atoms and cavity, cavity excitations decay quickly and the emission process resembles that of a light burst into free space. The density of states of the open cavity  leads to slight modifications in the transition rates between different spin states as compared to the emission into free space though. The corresponding dynamics of the collective spin after elimination of the cavity degree of freedom has been studied in depth previously \cite{braun:1998}. However, strong damping implies that the thermodynamics of the composite system may be strongly modified by the system-bath interaction so that no conclusions can be drawn from results found for the isolated system (e.g., the existence of a phase transition). In the opposite regime of vanishing dissipation,  the dynamics of the Dicke model has been studied in the past with the focus on  collapse-revival features, regularity, and quantum chaos \cite{altland:2012,*alvermann:2012,*bakemeier:2013}. This may 
give insight into collective properties of the compound but does not allow to access stationary state features for longer times.

In the following, we thus concentrate on finite but weak damping, where the dominant dissipative mechanism is due to a large but finite quality factor of the cavity. In actual realizations of the Dicke model, where trapped ensembles of cold atoms are brought into close proximity with a superconducting cavity, competing dissipation mechanisms like spontaneous emission, trap losses, processes involving additional states of a single atom, or dark states of the atomic ensemble are expected to be of minor relevance only, although they may have an impact in specific ranges of parameter space \cite{henschel:2010}. For circuit QED realizations with artificial atoms other noise sources must be considered, with decoherence most prominent in the Cooper pair boxes. We leave this issue for future work and,
in the spirit of a minimal setting, model the decay of electromagnetic excitations of the cavity in a standard manner by adding
to the Liouville-von Neumann equation for the density operator of the bare composite system a Lindblad-type dissipator. Hence, we write
for the time evolution of the reduced density operator of the cavity-atom aggregate
\begin{equation}
\label{EOM}
\frac{d}{dt} \hat \rho = \frac{1}{\rmi} \left[ \hat H _{\mathrm{Dicke}}, \hat \rho \right] + \mathcal L _{\mathrm{diss}} \left( \hat \rho \right)
\end{equation}
with the dissipator
\begin{equation}
\label{damping}
\hspace*{-1cm} \mathcal L _{\mathrm{diss}} \left( \hat \rho \right) = \kappa \left( \overline{n} + 1 \right) ( 2 \hat a \hat \rho \hat a ^{\dagger} - \hat a ^{\dagger} \hat a \hat \rho - \hat \rho \hat a ^{\dagger} \hat a ) \;
 +\; \kappa \overline{n} ( 2 \hat a ^{\dagger} \hat \rho \hat a - \hat a \hat a ^{\dagger} \hat \rho - \hat \rho \hat a \hat a ^{\dagger})\, . \quad
\end{equation}
Here, $\kappa$ is the damping rate and $\bar{n}$ is the thermal Bose occupation factor for an environmental photon with cavity energy $\omega$. Note, that we employ the Lindblad dissipators derived for radiation damping of the cavity without coupling to the spin. This
 captures  sufficiently accurately the non-equilibrium dynamics from the initial transient period up to long times where final equilibration sets in, and thus the regime where we expect the Dicke Hamiltonian to be realized in actual mesoscopic systems. Of course, for weak atom-field coupling, this modeling of dissipation is correct on all time scales.

\section{Dynamics of the cavity occupation}\label{cavoccu}
In this section the dissipative dynamics of the field mode or, equivalently of the cavity occupation according to (\ref{EOM}), is analyzed in detail on various time scales. The goal is to reveal the relation between and to characterize the nature of atom-field correlations during the initial and the final stage of the relaxation process.

\subsection{Emission burst and relaxation\label{burst}}
\begin{figure}[hp]
\centerline{\includegraphics[width=0.82\columnwidth]{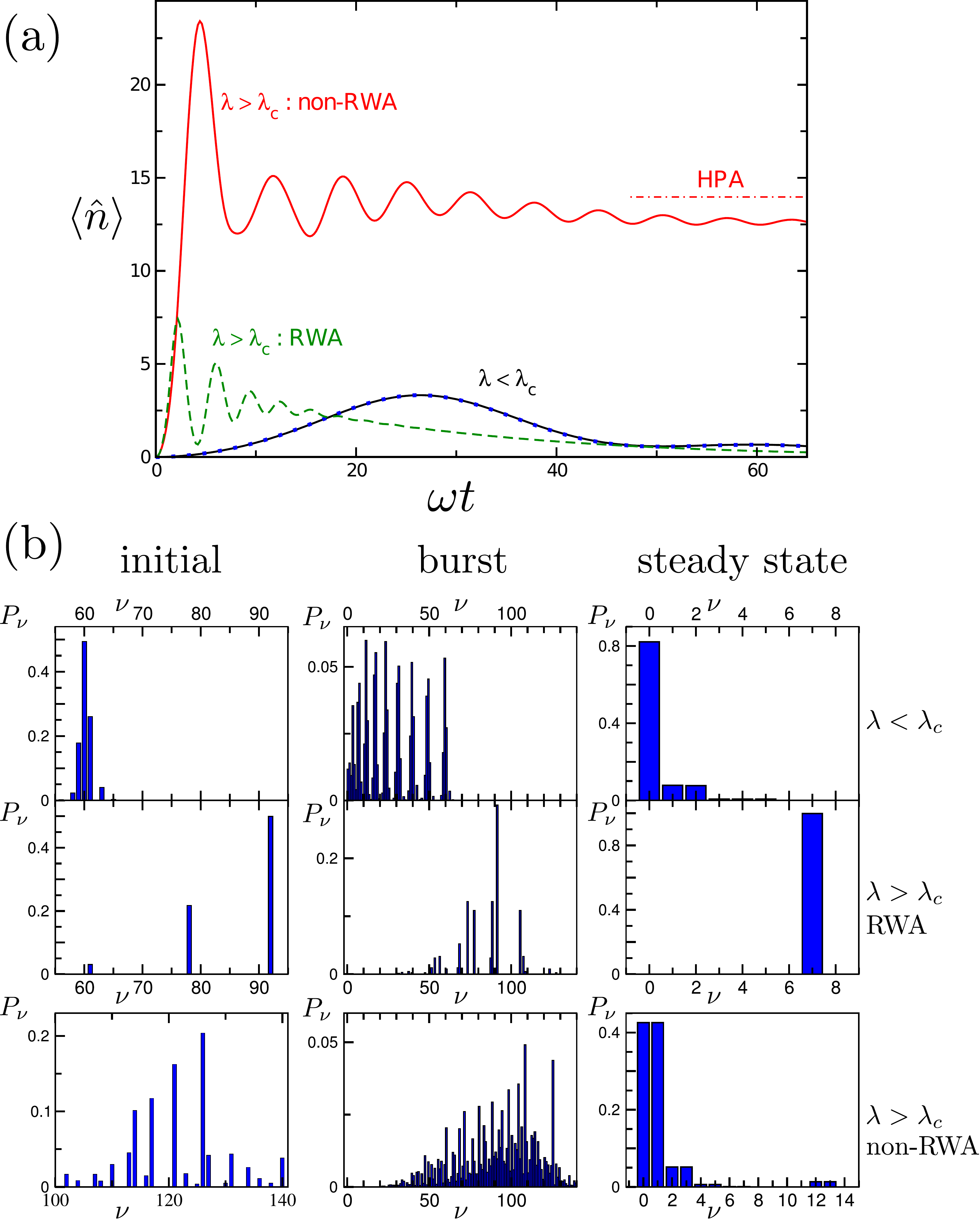}}
\caption{(a) Dynamics of the mean cavity occupation $\langle \hat{n}\rangle$ for coupling strength above/below the critical coupling and with/without the rotating wave approximation of the coupling term.
Above the critical coupling and without RWA the system equilibrates to steady state with finite mean cavity occupation (for comparison the HP result is shown). Employing RWA (green dashed and blue dotted lines) the system does not equilibrate to its ground state but reaches a steady state, where the cavity is empty. (parameters: $N=10,\;\kappa/\omega=1/20,\; \bar{n}=0$ and $\lambda=0.2\,\lambda_\mathrm{c}$ \--- normal, $\lambda=2.4\,\lambda_\mathrm{c}$ \--- superradiant)\\
(b) Occupation of eigenstates (arranged in order of ascending energy) for the initial state ($|n=0,\,M=+J\rangle$), an intermediate "burst" time (when $\langle \hat{n}\rangle$ is maximal), and in the long-time limit. At the burst many eigenstates are involved. The steady state occupation highlights the potential structure above/below the phase transition (see subsection~\ref{burst} in the main text). (parameters as above, but $\bar{n}=0.1$ for bottom right panel)
}
\label{general_dynamics_fig}
\end{figure}

 The situation is considered where initially the compound is prepared far from its equilibrium state with all atoms in the excited state and the cavity in its ground state.  We first concentrate on the zero temperature case and later on comment on finite temperature effects. As expected, the atom-cavity system starts to develop a coherent flow of excitations associated with an oscillatory pattern in typical observables such as  the mean cavity occupation $\langle \hat{n} \rangle$ (see \fref{general_dynamics_fig}) and  spin expectation values $\langle \hat{J}^{x/y/z} \rangle$ (not shown). Due to the leaky cavity these features are damped out so that the system finally relaxes to a stationary state. In \fref{general_dynamics_fig}(a) we contrast the dynamics of  $\langle \hat{n} \rangle$ for couplings $\lambda$ below and above  the critical coupling $\lambda_\mathrm{c}$. Results for the RWA model are included as well.
  In the regime below $\lambda_\mathrm{c}$, both the full Dicke model and its RWA version basically coincide and predict a decay towards 
  $\langle \hat{n} \rangle=0$. This is no longer the case in the regime above the critical coupling. There, the full model approaches asymptotically a finite cavity occupation in agreement with thermodynamic calculations (cf.~\cite{hepp:1973,emary:2003_2}) and (\ref{HP_meanvalues})) which cannot be captured by the  RWA dynamics.

Insight into this failure of the RWA model  can be obtained from the master-equation (\ref{EOM}). One easily finds in the stationary limit $\frac{\rmd}{\rmd t}   \langle \hat{n} \rangle|_{t\rightarrow \infty} = 0  = \frac{\rmd}{\rmd t} \langle \hat{J} ^z \rangle|_{t\rightarrow \infty}$ that
\begin{equation}
\label{n_limit}
\langle \hat n \rangle = \overline{n} + \rmi \frac{\lambda}{\sqrt{N} \kappa} \left( \langle \hat a \hat J ^- \rangle - \langle \hat a ^{\dagger} \hat J ^+ \rangle \right)\; ,
\end{equation}
where the latter counter-rotating terms are absent if the RWA is employed.
Hence, within RWA the Lindblad-damping term used in (\ref{damping}) always drives the system into a the thermal state of the bare cavity independent of the cavity-atom coupling. Consequently, no signatures of the phase transition can be found in the dynamics of the cavity occupation for the RWA model. In contrast, in the full model $\left( \langle \hat a \hat J ^- \rangle - \langle \hat a ^{\dagger} \hat J ^+ \rangle \right)$ plays the role of an order parameter which is zero below and takes finite values above the phase transition.
Dissipation guarantees that the compound relaxes to the correct correlated thermal state which  approaches with increasing $N$ and for $T=0$ the result known from the HP treatment. Of course, for any finite $N$ the phase transition (for example $\langle \hat{n} \rangle_{t \rightarrow \infty} vs.\ \lambda$)  is smeared out.

This crucial importance of the interplay of non-RWA terms and the particular damping mechanism for the relaxation towards the correct ground state has not been fully appreciated  in previous studies of dissipative Dicke models \cite{dimer:2007,henschel:2010}. Note though that in~\cite{henschel:2010} the RWA together with dissipation induced by atomic losses also yields signatures of a phase transition in the dynamics. 

The benefit of the RWA is that the set of equations of motions for $\langle \hat{n} \rangle$ and $\langle \hat{J}^z \rangle$ close in the stationary limit. This is no longer the case for the full model. Analytical progress can be made, however, for the short-time dynamics starting from the initial state of fully excited spins. One finds for the initial buildup of cavity excitations in the weak damping limit $\langle \hat{n} \rangle = \lambda^2 \;t^2$ (cf.~\fref{burst_fig}) independent of the number of atoms $N$ and valid for both the full model and for the RWA.

To elucidate further details of the dynamics and particularly make progress in understanding the nature of the first emission peak
in \fref{general_dynamics_fig}(a), it is instructive to analyze how different eigenstates $| \nu \rangle$ of the Dicke Hamiltonian participate during different stages of the dissipative time evolution. In \fref{general_dynamics_fig}(b)  probabilities $P_\nu$ to find the system in eigenstate $| \nu \rangle$ at three different stages of the dynamics are displayed: at $t=0$ (left), at the "burst time", when the first peak in $\langle \hat{n} \rangle$ occurs (middle), and for long times (right) for couplings below and above the critical coupling (RWA data are included as well \footnote{Far below the critical coupling the probability distribution with and without RWA is obviously very similar.}).

While in the normal regime the low energy sector is dominantly occupied throughout the time evolution, this changes drastically for $\lambda>\lambda_\mathrm{c}$, particularly for times around the first emission peak. 
There, a large number of eigenstates participate substantially.
Asymptotically, the probabilities $P_{\nu}$ show a near-degeneracy of pairs of eigenstates which is reminiscent of the exact degeneracy in the thermodynamic limit according to the HP treatment. As discussed above, finite size effects induce correction terms to the HP Hamiltonians $\hat{H}^\pm$ and lift their degeneracy. How many pairs of near-degenerate eigenstates occur, is a measure of the height of the barrier between the potential minima at $(\pm\sqrt{2\alpha}, \mp \sqrt{2\beta})$. To illustrate the appearance of these near-degenerate states, in \fref{general_dynamics_fig}(b) data are shown at a slightly elevated temperature $\bar{n}=0.1$. Note that one has also finite occupations in some higher lying eigenstates due to the small but finite value of the damping parameter $\kappa$.
Apparently, the data for the RWA model strongly deviate from these findings for $\lambda>\lambda_\mathrm{c}$.
In particular, in the stationary limit the system is not relaxed to the lowest energy eigenstate but to a state close to the ground state of the uncoupled system.

\subsection{Characteristics of the emission burst: Is it superradiance?}
In this subsection we will argue that the first pronounced peak in the cavity occupation of the damped  Dicke model carries signatures of the superradiant free\--space burst and reflects its characteristic features. These are the crossover in the time-dependence of the emission rate from an exponential decay for a single atom to a peak, where for larger $N$ the height of the peak grows as $N^2$, its temporal width decreases as $N^{-1}$ and the integrated emission is proportional to the total initial excitation energy and, hence, to $N$.

\begin{figure}[ht]
\centerline{\includegraphics[width=0.95\columnwidth]{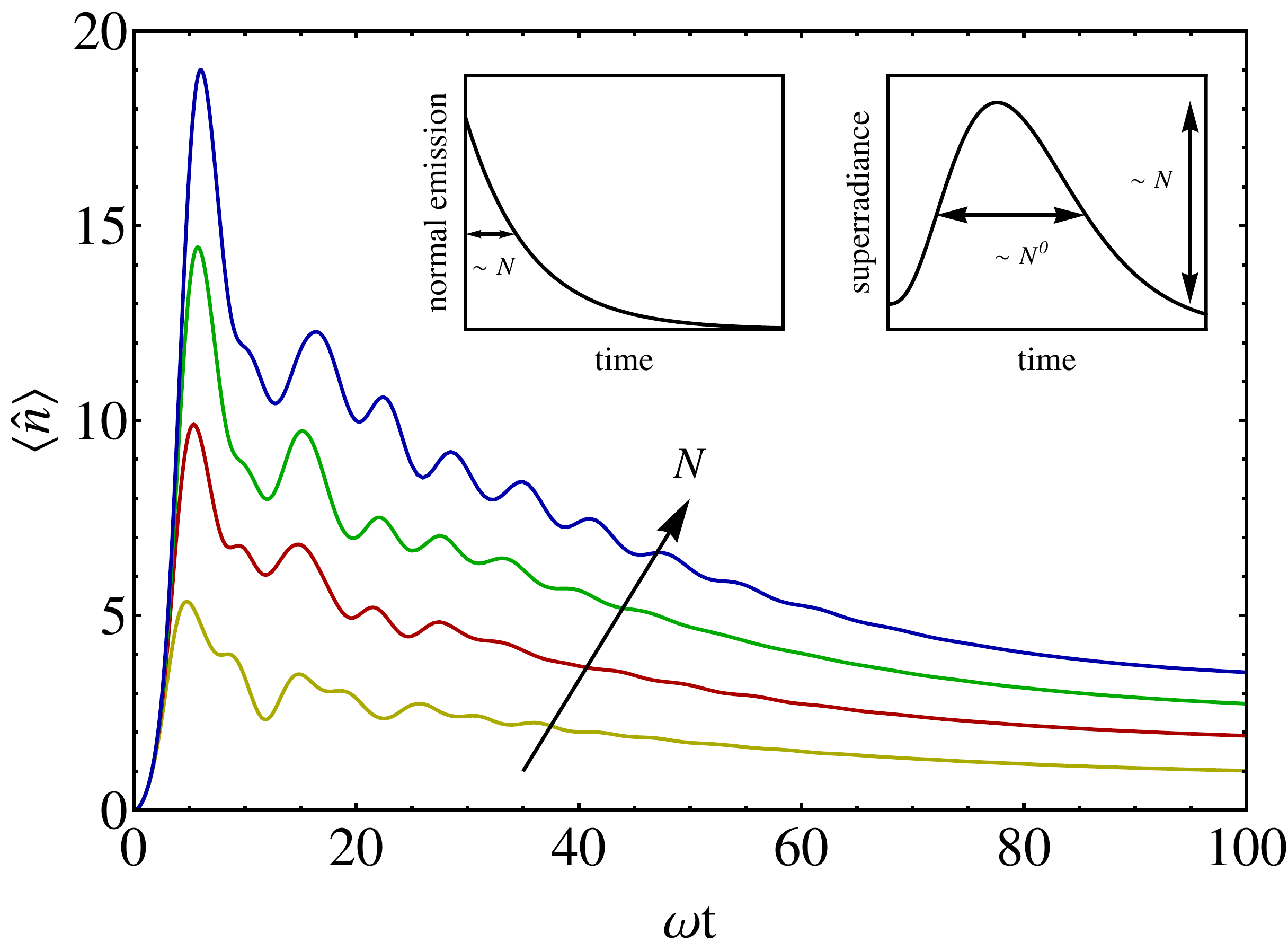}}
\caption{ Dynamics of the mean cavity occupation $\langle \hat{n}\rangle$ of the superradiant damped Dicke model ($\lambda=1.2\,\lambda_\mathrm{c},\;\kappa/\omega=1/40$) for increasing number of atoms, $N=5,\,10,\,15,\,20$.\\
The insets contrast the emission of light into free space from $N$ independent atoms (normal) and collective emission (superradiant) where the single-atom to field coupling is rescaled $g\sim1/\sqrt{N}$ as in the Dicke case.
 The first peak in $\langle \hat{n}\rangle$ has the same scaling characteristic as the superradiant  burst into free space: the peak height increases linearly with $N$, the width is constant.}
\label{burst_fig}
\end{figure}

These scaling properties are modified when one considers a fixed value of the parameter $\lambda$ upon increasing $N$, the typical
scenario for the analysis of the phase transition in the Dicke model. 
Then,  the rescaled coupling constant between a single atomic excitation and the e-m field is $g=\lambda/\sqrt{N}$ so that
 the emission rate due to a single excitation is reduced $\propto g^2=\lambda^2/N$ for a larger number of atoms. Accordingly,
  the characteristics of the light emission are as follows (see \fref{burst_fig}): 
(i) For  $N$ independent
atoms the radiation follows an exponential decay from a fixed ($N$-independent) value over a timescale growing linearly
with $N$; (ii) For a superradiant burst of $N$ cooperating atoms a peak with height $\propto N$ and constant width $\propto N^0$ emerges.\\

We can now test these signatures for the mesoscopic dissipative Dicke model (\fref{burst_fig}). Indeed, for $\lambda>\lambda_\mathrm{c}$
we find an increase of the height of the first peak with growing $N$, while its width remains basically constant. Quantitatively,
for the $N$-dependence of peak height and width (defined here as the left-sided half width at half maximum value) the expected scaling
with $N$ and $N^0$, respectively, is obtained (not shown).
While for the results discussed so far we considered the resonant case, where the cavity energy equals the energy needed for atomic excitations, we numerically checked also the off-resonant case. It turns out that the discussed dynamical features are not very sensitive to weak de-tunings. This is in stark contrast to the purely coherent dynamics, where complex collapse-revival dynamics with domains of regularity and quantum chaos occur \cite{altland:2012,*alvermann:2012,*bakemeier:2013}. We enter this dynamical range for extremely weak dissipation as well.

In conclusion, our findings strongly suggest that the physics of the free-space superradiant burst can indeed be recovered in the damped dynamics of the Dicke model for suitable parameters. In an actual experimental realization with a cold-atom cloud coupled to a superconducting stripline-cavity along the lines of~\cite{bernon:2013,*bothner:2013}, the atom number $N$ is orders of magnitude larger than values which can be realized in full numerical simulations. Nevertheless, the scaling characteristics of the peak will allow for a wide range of damping rates, where friction is strong enough to suppress coherent collapse-revival dynamics but weak enough to achieve prominent radiation bursts with peak cavity occupations $\gg 1$. 

\section{Monitoring collective dynamics}\label{monitor}
\subsection{Dicke dynamics in phase space}
Our discussion of the dynamics of the dissipative Dicke model has so far been focused on a single observable, the mean cavity occupation $\langle \hat{n} \rangle$, which is certainly not the only way, to gain experimental insight into the dynamics. A more complete characterization of the cavity dynamics is offered for example by the (reduced) Wigner density of the cavity degree of freedom, which is routinely measured for stripline resonator systems (cf.~\cite{kirchmair:2013} for a recent example).

In particular, in the thermodynamic limit $N\rightarrow \infty$ of a nearly classical spin, we may expect the Wigner density, which has a simple classical interpretation, to be exceptionally suitable. One question of interest is to distinguish sub\-- and superradiant coupling without waiting for the relaxation of the system to its stationary state. That way, such a distinction would be immune to the danger that the long-time limit is influenced or covered by extraneous terms, not present in the Dicke Hamiltonian (e.g., non-homogenous coupling \cite{braun:2011,zou:2014}) or additional or different damping mechanism (e.g., atom decay \cite{henschel:2010}). In fact, we find that the Wigner density dynamics shows clear signatures, reflecting the contrast between a unique mean field solution for the ground state below the critical coupling and two equivalent solutions above, already for short times and a small number of atoms.

\begin{figure}[t]
\centerline{\includegraphics[width=.95\columnwidth]{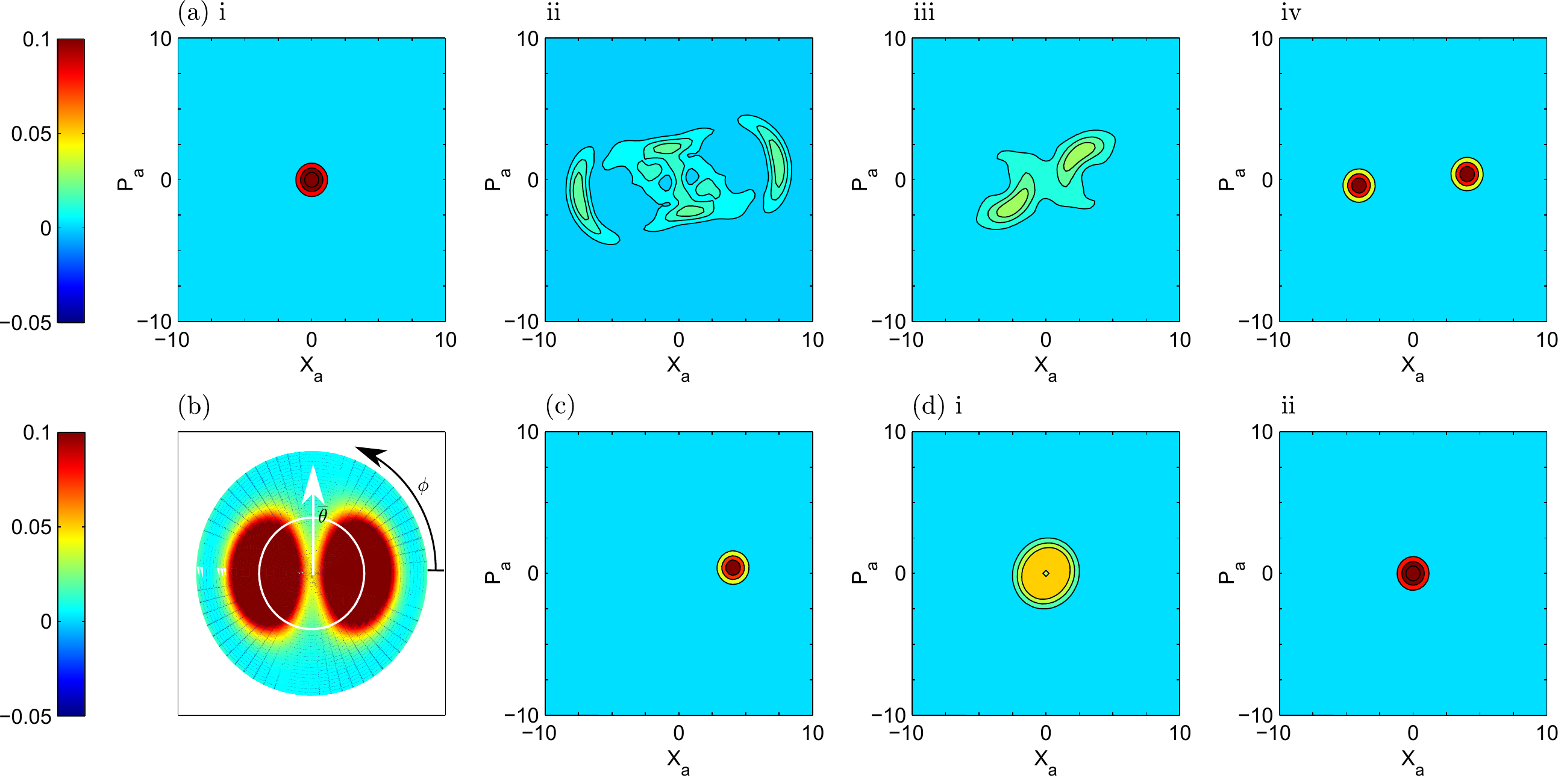}}
\caption{Dynamics in phase space.\\
(a) Reduced Wigner density of the cavity above the critical coupling ($\lambda=1.2\lambda_\mathrm{c}$): Starting from a ground state, i, the spin system excites the cavity. Occupation oscillates around, ii at $t= t_\mathrm{burst}$ and iii at $t=2t_\mathrm{burst}$, and finally relaxes to two minima positions iv.\\
(b) shows the Husimi function 
of the spin degree of freedom in the stationary state, which shows a similar relaxation towards two final peaks. The spin is then directed towards two opposite points on the southern hemisphere (cf.~the equator indicated by the white circle, while the south pole ($\bar{\theta}=\pi-\theta\equiv 0$, corresponding to $|M=-J\rangle$) is in the center of the plot.\\
(c) shows, that the spin-reduced $M_x=-J$ component of the cavity Wigner density relaxes to one of the stationary state positions.\\
Below the critical coupling ($\lambda=0.2\lambda_\mathrm{c}$) the cavity Wigner density starting from the ground state, cf.~(a) i, undergoes a breathing-type oscillation, (d) i at $t=t_\mathrm{burst}$, and finally relaxes to a final peak close to the ground state, (d) ii. (other parameters: $N= 6,\;\kappa/\omega=1/10,\;\bar{n}=0$)
}
\label{Wigner_fig}
\end{figure}

In \fref{Wigner_fig}(a) we show the reduced Wigner density of the cavity (in dimensionless coordinates $(X_a, P_a)$) well above the critical coupling at various times. The Wigner function of the initial state at $t=0$ is a Gaussian peak centered at $(X_a=0,\,P_a=0)$. It starts to spread out in two arms in the positive and negative $X_a$ direction and evolves into the structure shown in \fref{Wigner_fig}(a) ii) at the time of the burst, $t=t_\mathrm{burst}$. A rich fine structure of complex interference patterns remains for some time, while substantial weight is assembling in two peaks (cf. (a) iii at $t=2\,t_\mathrm{burst}$). These peaks rotate around and finally slowly relax towards two final positions in phase space (cf. (a) iv), which correspond to the two equivalent solutions of the HP approximation in the $N\rightarrow \infty$ limit.

The spin-dynamics can similarly be visualized by another quasi probability-distribution, the Husimi function, relying on the concept of spin-coherent states. We show the final state's reduced spin-Husimi function in the phase space spanned by angular variables $(\phi, \theta)$, which indicate the "direction" of the spin vector, in \fref{Wigner_fig}(b) for the same superradiant coupling as in (a). As for the cavity degree of freedom, we find a two peak structure for spin values corresponding to the two equivalent solutions of the HP result, cf.~(\ref{HP_meanvalues}).

For subcritical coupling the initial central Gaussian peak broadens [see \fref{Wigner_fig}(d) i for $\lambda=0.2\,\lambda_\mathrm{c}$ and $t=t_\mathrm{burst}$] and re-sharpens in a breathing-type oscillation, which is damped out to relax to a final Gaussian peak centered around the unique minimum $(0,0)$ [see \fref{Wigner_fig}(d) ii].

The long-time behavior observable in \fref{Wigner_fig}(a) iii, iv and (b) can be intuitively understood in the HP picture as damped dynamics well within the two equivalent parabolic potentials in cavity and spin variables. For the short-time behavior, however, the HP approximation is not applicable. To explain the observed splitting of the central peak into two arms [\fref{Wigner_fig}(a) ii], one can employ the representation of the Hamiltonian in a $\hat{J}^x$ basis yielding the shifted coupled parabolas of equation~(\ref{eq:Dicke-Hamiltonian-Larson}). Starting in the initial state, which is not a $\hat{J}^x$ eigenstate, each $\hat{J}^x$ component (which \-- as a cavity ground state \-- is a centered Gaussian in $x$-space) evolves in its correspondingly shifted parabola and starts to move into positive or negative $x$-direction. According to this picture, we expect, e.g., the $M_x=\pm J$ component to start to move into positive/negative $x$-direction. Indeed, this behavior is observed in the corresponding spin-projected Wigner density of the cavity in \fref{Wigner_fig}(c) [same coupling and time as in (a) ii]. This argument has so far neglected the coupling between the motion in the various parabolas induced by the $\hat{J}^z$ term in (\ref{eq:Dicke-Hamiltonian-Larson}), which causes transitions and interference structures, which become prominent, when the various component-wavepackets meet at a crossing of the parabolas.

\subsection{Indicators of sub/superradiant coupling}
The concept of a phase transition applies to a system in the thermodynamic limit ($N\rightarrow\infty$), which after a long time ($t\rightarrow\infty$) is driven to equilibrium by a vanishingly weak coupling to an environment ($\kappa \rightarrow 0$). For that case, the mean cavity occupation, $\langle \hat{n} \rangle/J$, is an order parameter for the Dicke model, i.e., an observation of its value clearly indicates whether the coupling is sub\-- or superradiant, $\lambda \gtrless \lambda_\mathrm{c}$. In our case, however, we are interested in an observable distinguishing between the weak and strong coupling case for a mesoscopic ($N\gtrsim 10\--100$) Dicke aggregate with finite coupling to the environment on short to intermediate time scales.
\begin{figure}[t]
\centerline{\includegraphics[width=.95\columnwidth]{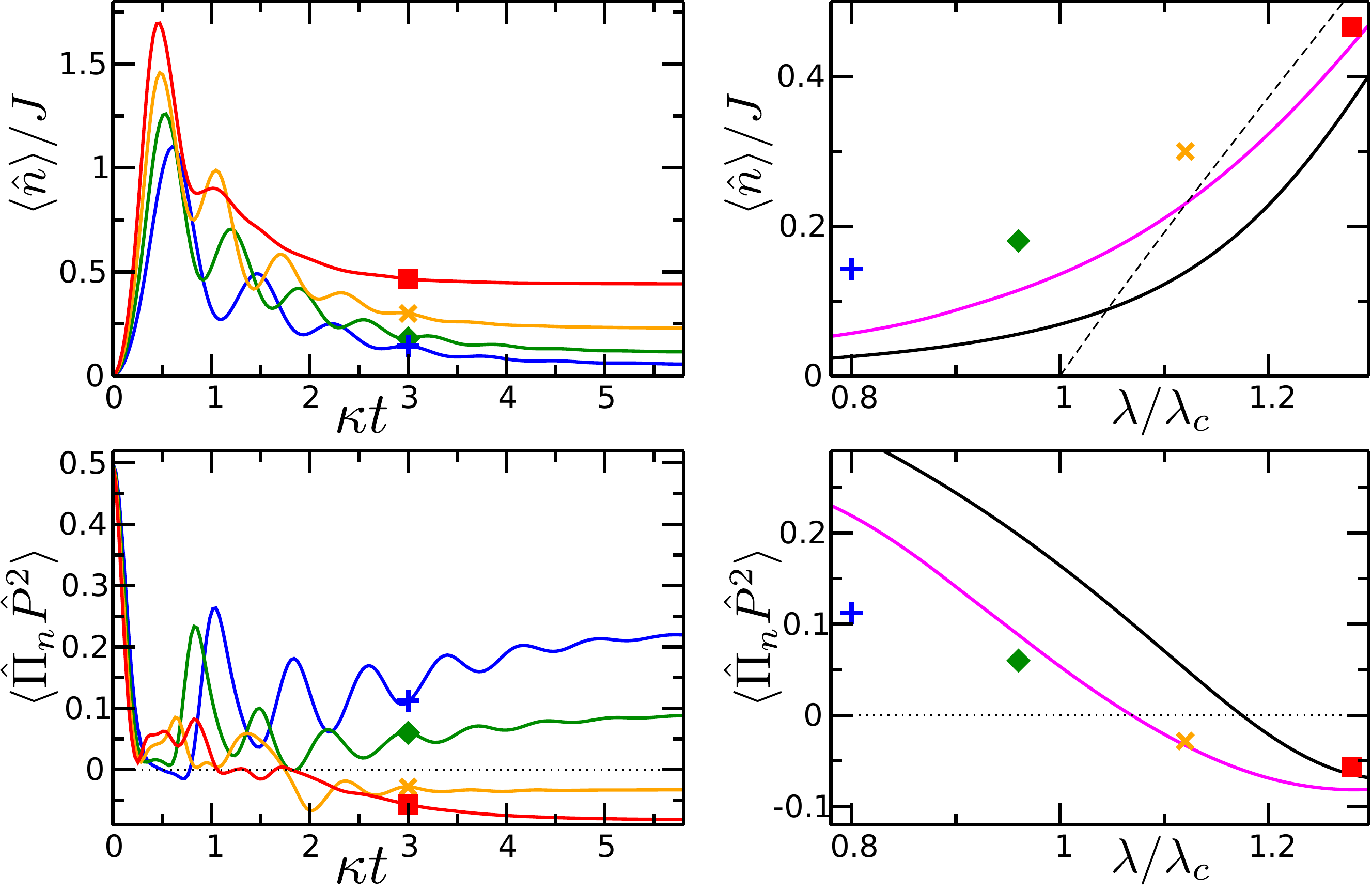}}
\caption{Observables as indicator of normal or superradiant phase.\\
The order parameter, $\langle \hat{n}\rangle/J$, of the thermodynamic phase transition does not indicate clearly, if the coupling 
$\lambda$ is above or below the critical value $\lambda_\mathrm{c}$, when the dynamics of the damped (mesoscopic, $N=6$) Dicke model
 is observed on short time-scales only (upper row). \\
Other observables, like the curvature of the (cavity-)Wigner density, $\langle \hat{\Pi}_n\;\hat{P}^2 \rangle$, which quickly goes
 to positive (negative) values for coupling values below (above) the critical value, can give clearer signatures (lower row).\\
The left column shows the dynamics of the two observables for various values of the coupling ($\textcolor{blue}{\pmb{+}}:\lambda/\lambda_\mathrm{c}=0.8,\;
\textcolor{Green}{\blacklozenge }:\lambda/\lambda_\mathrm{c}=0.96,\;
\textcolor{Orange}{\pmb{\times }}:\lambda/\lambda_\mathrm{c}=1.12,\;
\textcolor{red}{\blacksquare}:\lambda/\lambda_\mathrm{c}=1.28,\;
$) for times on the order of the inverse damping rate, $\kappa t \sim 1$.\\
In the right column the value of the observables at an intermediate measurement time, $\kappa t =3$ (symbols), is contrasted
 to its stationary limit with ($\kappa/\omega=1/10$ \-- magenta), and without damping (black solid line). The order parameter 
 observable, $\langle \hat{n}\rangle/J$ (upper right panel), while distinctive in thermodynamic limit (black dashes), is less suited 
 as an indicator of super/subradiant coupling, than other observables, like the Wigner curvature (lower right panel), which shows a 
 distinctive sign change at (approximately) the critical coupling strength.
}
\label{newobs_fig}
\end{figure}
The latter requirement is due to the presence of additional physical processes in any actual experimental realization, which our description 
of the dynamics of the damped Dicke problem does not take properly into account as discussed in the Introduction.

It is, therefore, advisable to search for alternative observables, which may in this scenario be better indicators of sub\-- or superradiant coupling than the cavity occupation.
The investigation of the dynamics of the Wigner density has revealed one distinctive difference between the sub\-- and superradiant case already apparent after short times: namely, a substantial fragmentation of the Wigner density into two parts, oscillating and finally relaxing to the two equivalent minima of the superradiant case. This is, for instance, reflected in the curvature of the Wigner density (in $X_a$-direction)
\[
\frac{\rmd^2}{\rmd X_a^2} \left.W(X_a,P_a)\right|_{X_a=0=P_a} = -\frac{4}{\pi} \langle \hat{\Pi}_n\;\hat{P}_a^2 \rangle
\]
at the center of the  cavity phase space $X_a=0=P_a$, where  the second expression containing the parity operator for the cavity degree of freedom, $\hat{\Pi}_n$, follows from the definition of the Wigner density.

Note, that this observable, which can for example be gained from a subset of measurements of the full Wigner density, is but one example of a higher order correlation function of cavity operators. Other combinations, also mixed cavity-spin correlators may be of similar or in other parameter regimes even of superior use for deducing the coupling strength from observations of the short-time dynamics. In \fref{newobs_fig}, we demonstrate that the sign of the new observable can indeed give a clear indication of sub\-- or superradiant coupling strength after a short observation time of the order of the inverse damping rate, e.g., $\kappa t \le 3$ (lower row). In contrast, the dynamics of the order parameter of the thermodynamic case, $\langle \hat{n} \rangle/J$, up to this time only does not easily allow to gauge, how close to relaxation the system is (upper left panel of \fref{newobs_fig}); even less does it show any distinguishing features around the transition from sub\-- to superradiant coupling (symbols in upper right panel).

\section{Conclusion}
Investigating the dissipative Dicke dynamics in a system where a mesoscopic cold atom cloud is coupled to a superconducting cavity, we found that two facets of superradiance physics appear. In the long term, dissipation drives the relaxation of the system towards its equilibrium state, which shows the profoundly different properties of a normal and a superradiant phase below and above the critical coupling strength. On short time scales, a peak in the cavity occupations was shown to exhibit the characteristic scaling properties of the free-space superradiant radiation burst.
The radiation burst is further characterized by the participation of many eigenstates in the dynamical evolution for which non-RWA terms are of crucial importance.

To overcome issues problematic for observing the long-time limit of full relaxation, we investigated the phase-space dynamics of cavity and the atomic degree of freedom, where the reduced (quasi-)probability distributions were found to reflect the appearance of two degenerate minima in the superradiant regime. This allows to identify suitable obervables and features in their transient dynamics, which allow a clear distinction of sub\-- and superrradiant coupling on short time scales.

\section*{Acknowledgments}

We acknowledge fruitful discussions with C. Bokas, J. Fort\'{a}gh, H. Hattermann, and R. Kleiner. S.\,F., J.\,A., and B.\,K. thank for the kind hospitality
of the Department of Physics and Astronomy at Dartmouth College. Financial support was provided by the Harris-Foundation through a
distinguished visiting professorship at Dartmouth College (J.A.), the DAAD PROMOS program (S.F.),  and the German Science Foundation through SFB/TRR 21.


\providecommand{\newblock}{}

\end{document}